\documentclass[11pt]{article}

\usepackage[a4paper,lmargin=2.5cm,rmargin=2cm,tmargin=1cm,bmargin=1cm,includehead,includefoot]{geometry}

\usepackage{graphicx,amsmath,url}
\usepackage{microtype,lmodern}
\usepackage{mathtools,amssymb,gensymb}
\usepackage{tgtermes}
\usepackage[T1]{fontenc}
\usepackage[utf8]{inputenc}

\usepackage{acronym}
\usepackage{amsmath}
\usepackage{array}
\usepackage{bbold}
\usepackage{booktabs}
\usepackage{dsfont}
\usepackage{fancyhdr}
\usepackage{relsize}
\usepackage[font={small,up,singlespacing}]{subcaption}
\usepackage[backend=bibtex,style=chem-angew,sorting=none,autocite=superscript]{biblatex}
\usepackage{pdfpages}
\usepackage{bbm}
\usepackage{color}
\usepackage{siunitx}
\usepackage{braket}
\usepackage{mwe}
\usepackage{biblatex}
\usepackage{multirow}
\usepackage{mathtools}
\usepackage{xcolor}
\usepackage{float}
\usepackage{colortbl}
\usepackage{siunitx}\DeclareSIUnit\molar{\textsc{M}}

\usepackage[colorlinks,breaklinks]{hyperref} 
\usepackage{cleveref}

\hypersetup{hypertexnames=true, linkcolor=blue, anchorcolor=black, citecolor=blue, urlcolor=blue}
\graphicspath{{./}{imglocal/}{img/}}
\title{\textbf{DASH: Dynamic Attention-Based Substructure Hierarchy for Partial Charge Assignment}}

\author{Marc T. Lehner,$^{a,\dagger}$ Paul Katzberger,$^{a,\dagger}$ Niels Maeder,$^a$ Carl C. G. Schiebroek,$^a$ Jakob Teetz,$^a$ \\ Gregory A. Landrum,$^a$ Sereina Riniker$^a$*}
\date{$^a$~Department of Chemistry and Applied Biosciences, ETH Zurich, Vladimir-Prelog-Weg 2, 8093 Zurich, Switzerland. *Email: sriniker@ethz.ch \\
$^\dagger$ These authors contributed equally}

\begin{document}

\maketitle

\section*{Abstract}

We present a robust and computationally efficient approach for assigning partial charges of atoms in molecules. The method is based on a hierarchical tree constructed from attention values extracted from a graph neural network (GNN), which was trained to predict atomic partial charges from accurate quantum-mechanical (QM) calculations. The resulting dynamic attention-based substructure hierarchy (DASH) approach provides fast assignment of partial charges with the same accuracy as the GNN itself, is software-independent, and can easily be integrated in existing parametrization pipelines as shown for the Open force field (OpenFF). The implementation of the DASH workflow, the final DASH tree, and the training set are available as open source / open data from public repositories. 

\section{Introduction}
Molecular dynamics (MD) simulations enable the time-resolved study of molecular systems and are therefore widely used in biology, chemistry, and material science. The physical interactions between the particles in the system are thereby approximated by a set of potential-energy functions (i.e., the force field) \cite{vanGunsteren2013_bioSimFF,ponder2003ff_proteinSim,mackerell2004empiricalFF,jorgensen2005eFunc_FF,vanGunsteren2012_35yearsFF,Riniker2018}. Applying Newton's equation of motion allows the propagation of the system through time. 
The quality of the MD simulations is determined by the approximations made in the functional form of the force-field terms as well as their parameters. 
For biomolecular simulations, fixed-charge atomistic force fields are predominantly used due to their reasonable accuracy and low computational cost \cite{Riniker2018}. In such force fields, the contributions are split into bonded terms (i.e., bond-stretching, bond-angle-bending, and dihedral-angle torsion) and non-bonded terms (i.e., electrostatic and van der Waals) \cite{Riniker2018}. 
The non-bonded terms describe the intermolecular interactions, which are directly related to experimental observables such as the density or the heat of vaporization of a compound. The calculation of the non-bonded interactions between the atoms in the system constitute the computationally most expensive part of every classical MD simulation. While the van der Waals forces decay quickly with increasing distance between the atoms, the long-range contribution of the electrostatic forces is non-negligible and many schemes have been developed for their efficient treatment (e.g., Ewald summation \cite{Ewald1921} based methods such as smooth particle mesh Ewald \cite{smoothPME}, or reaction-field (RF) correction \cite{Barker1973_ReactionField}). The slow decay of the electrostatic forces also means that small changes in the parameters (i.e. partial charges, dielectric constants) can lead to large changes in the potential energy.

Molecules can only have integer charges, but even in a simple Lewis representation the assignment of atomic formal charges can be ambiguous since they are not experimentally measurable and resonance structures can exist. Nevertheless, many techniques have been developed over the past decades to determine atomic partial charges, which can be used to predict chemical reactivities or perform MD simulations, among other applications. Early examples of such models include Gasteiger charges \cite{Gasteiger19783181_atomiCharge}, Hirshfeld type charges \cite{Hirshfeld1977}, MMFF charges \cite{Halgren1996_Mmff_og}, and Mulliken type charges \cite{Jakalian2002}. The partial charges are extracted from a quantum-mechanical (QM) calculation (e.g., Hartree-Fock (HF), density functional theory (DFT), or semi-empirical methods) and/or fitted to reproduce experimental properties. 
Mulliken-type charges, for instance, are calculated by integrating the electron density over the volume of the atoms. One of the most commonly used representative from this family are AM1 population charges, which employ the semi-empirical method AM1 \cite{Dewar1985_AM1} for the QM calculation. Additional bond charge corrections (BCC) are then applied to better reproduce the electrostatic potential (ESP) calculated with the more accurate HF method \cite{Dewar1985_AM1}. The resulting AM1-BCC model is a reasonably fast and reliable method, which is used in classical force fields such as the general AMBER force field(GAFF) \cite{wang2004gaff} and OpenFF \cite{openff_parsley}.
Recently, progress has been made with Hirshfeld type charges such as DDEC \cite{Manz2015_ddec} and MBIS \cite{Verstraelen2016_mbis}, showing that they are more accurate in reproducing ESP surfaces than AM1-BCC charges. However, the higher accuracy comes with increased computational costs. 
Additionally, the accuracy depends on the level of theory and basis set used in the underlying QM calculation, introducing hard limits on the feasibility for larger molecules like proteins. The computational time needed to extract partial charges also matters if the number and/or size of molecules is large, like in enzyme screens, where non-bonded interactions are used as features for substrate prediction \cite{mou2021_bioMolScans}, or for large virtual screening runs in drug discovery \cite{Wan2004_QSAR}.

Since accurate Hirshfeld type charges are computationally expensive and scale poorly with the number of atoms, alternatives based on machine learning (ML) have been explored in recent years. These attempts range from simpler regression \cite{rai2013charge_prediction_regression_ML_charges} or random forest models \cite{Bleiziffer2018} to more complex graph neural networks \cite{martin2019contradrg_ML_charges,kato2020ML_charges,Behler2021_HDNNPs,Jiang2022_GNN_partial_charge,Gallegos2022_NNAIMQ,wang2023espaloma_ML_charge}. Bleiziffer \textit{et al.} \cite{Bleiziffer2018} showed that a random forest model trained on DDEC partial charges (TPSSh/def2-TZVP level of theory with an implicit solvent with a dielectric permittivity $\epsilon = 4$) from 130'000 molecules could predict partial charges of unseen molecules reasonably well with an RMSE of 0.03\,e. More recent approaches have explored the usage of different charge models as well as other ML techniques \cite{Behler2021_HDNNPs,Jiang2022_GNN_partial_charge,Gallegos2022_NNAIMQ}. All of these ML approaches predict partial charges with good accuracy while offering a drastic increase in speed relative to performing a separate QM calculation for each new molecule. However, the ML models are generally not interpretable, there is a risk of overfitting, and most models do not provide uncertainties with their predictions. In addition, these models are highly dependent on the correct featurizers and library versions, which often do not have long-term stability in the rapidly evolving field of machine learning.

In an attempt to peer into the black box of ML models, explainable artificial intelligence (AI) tools have been developed (e.g., LIME \cite{Ribeiro2016_LIME_explainer} or SHAP \cite{Lundberg2017_shap_explainer}) to explain the predictions of a model in a retrospective manner. These tools are often based on the idea of local linear approximations. A recent addition is the GNNExplainer \cite{Ying_GNNExplainer} for graph neural networks (GNNs), where each neighbor of a certain atom is assigned an attention value, representing the importance that this neighbor has in the prediction of the value for the given atom. There are many different ways to get such an attention score. GNNExplainer is a stochastic explainer, randomly generating subgraphs and comparing the predictions of the model on these subgraphs to the predictions on the full graph. This provides the advantage that the approach is agnostic to the architecture of the model.
While such explanation-based methods are able to explain a specific prediction and assign a measurement of importance to each neighboring atom, they are also computationally expensive due to the iterative and stochastic learning of the method, presenting a challenge for large data sets. 

In this work, we train a GNN on a substantially increased data set of QM reference partial charges compared to Ref.~\cite{Bleiziffer2018}. We demonstrate that the attention values extracted from this GNN model are in agreement with common chemical knowledge. Unlike chemical intuition, however, the attention values are quantitative, enabling us to rank certain atoms and functional groups over others. Thus, we can not only extract the important features as patterns but also use the attention values to construct a dynamic hierarchical tree structure to assign partial charges without the GNN model. The resulting dynamic attention-based substructure hierarchy (DASH) is independent of the ML software library with which the model was built and provides accuracy similar to the underlying GNN. Moreover, the DASH is human-readable and provides confidence values for each result.

\section{Methods}

\subsection{Data Set Generation}
A generally applicable force field for organic molecules needs to cover a large chemical space, including the combination of functional groups. 
In Ref.~\cite{Bleiziffer2018}, we generated a data set with a large coverage of chemical space while focusing on lead-like compounds (molecular weight in the range 250 - 350 g/mol), such that the molecules were small enough for high-level QM methods. We used the unique bits of Morgan fingerprints with a radius of 2 (MFP2) of all lead-like compounds in ChEMBL \cite{Bento2014_CHEMBL} and ZINC \cite{Sterling2015_ZINC} to select a diverse subset of 130'000 molecules that represented all MFP2 bits. At the time, we considered only one conformer per molecule since the conformational dependence of DDEC partial charges was found to be low overall. However, this can introduce noise for molecules for which the conformational dependency of the charge assignment is above average. This is, for instance, the case for molecules that are symmetric in the 2D graph but asymmetric in the 3D conformation. In the recently published QMugs data set \cite{Isert2022_qmugs} three conformers were included for each of the 200'000 molecules, using a semi-empirical method for geometry optimization and DFT for the calculation of the QM properties.

\subsubsection{Selection of Molecules}
In this work, we generated an extended data set by collecting and filtering molecules from four different sources:
(i) the QMugs data set \cite{Isert2022_qmugs},
(ii) the training set from Ref.~\cite{Bleiziffer2018}, 
(iii) lead-like molecule from ChEMBL version 30 (filtered as in Ref.~\cite{Bleiziffer2018}),
and (iv) organic liquids from Refs.~\cite{caleman2012vanDerSpoel_data,horta2016gromos2016H66_organic_liquids,schuler2000gromos_alkanes,moine2017estimation}. 
The goal was to have the minimal number of molecules that represent all unique MFP2 atom environments found in the lead-like compounds of ChEMBL at least five times.

First, the QMugs data set was filtered by removing larger molecules (molecular weight $>$ 500 g/mol), which are impractical for high-level QM calculations, and by iteratively removing molecules if the bits of their MFP2 fingerprint were already represented at least five times by the other molecules in the data set. Note that the molecules in the QMugs data set have a neutral formal charge but molecules can be zwitterions, i.e., contain a positively and negatively charged functional group. With the QMugs subset at hand, molecules from the other sources were added iteratively if their MFP2 fingerprints contained new bits. For this, the molecules were sorted by the number of 'unseen bits' in their MFP2 fingerprint, and the list was updated after each addition of a molecule. 
Finally, we observed that the data set did not contain a diverse enough set of charged nitrogen environments (e.g. protonated amines), so 21 manually selected molecules with charged nitrogen-containing functional groups (but a net zero charge) were added.

The final data set contains 341'811 molecules with elements from the organic set (C, H, N, O, P, S, Cl, Br, I, F) and a molecular weight of up to 500 g/mol.

\subsubsection{Conformer Generation and Extraction of Atomic Partial Charges}
For the molecules originating from the QMugs data set, all three semi-empirically optimized conformers were considered. For the other molecules, three conformers were generated with a similar workflow as in Ref.~\cite{Isert2022_qmugs}. The ETKDG conformer generator as implemented in the RDKit \cite{Landrum2023_rdkit} was used to generate three diverse conformers. The three conformers of a compound were treated as separate molecules in the workflow -- except when splitting the data set into training and test sets, i.e., all conformers of a given molecule were always assigned to the same split. 
The conformers were first optimized with the MMFF94 force field \cite{Halgren1996_Mmff_og} as implemented in RDKit \cite{Tosco2014_MMFF94inRDKIT} for an initial relaxation. 
These conformers were further optimized with the semi-empirical method XTB-GFN2 \cite{XTBGFN2} for 100 cycles with an implicit solvent ($\epsilon=4.6$) using the software package PSI4 \cite{Turney2012_psi4}. The choice of this implicit solvent was based on Ref.~\cite{Bleiziffer2018}, where we showed that this dielectric permittivity leads to partial charges most compatible with the van der Waals parameters of existing force fields (compared to $\epsilon=1$ (vacuum) or $\epsilon=78$ (water)). A single point DFT calculation was performed for each optimized conformer with the TPSSh functional \cite{TPSSh} and a def2-tzvp basis set \cite{def2tzvp1,def2tzvp2} in PSI4. The PCM implicit solvent model was used with chloroform as implicit solvent. 
MBIS charges were calculated with the {\tt oeprop} function in PSI4 with the wave function from the single point TPSSh calculation, with at most 300 iterations, $10^{-4}$ as convergence value, 75 radial points and 302 spherical points.

\subsection{Training of the Graph Neural Network}

\subsubsection{Model Architecture}
The model architecture was based on the first two layer types of the Attentive FP network developed by Xiong \textit{et al.} \cite{Xiong2020_AttentiveFP} (i.e., the input layer and the attention layer for atom embedding) and a three-layer multi-layer perceptron (MLP) \cite{Rosenblatt1963_MLP} with ReLu activation functions \cite{Fukushima1969_relu}. The atomic features are first passed through the input layer followed by five layers of the attention layer for atom embedding type and are then decoded by the MLP. In a final step, the predicted partial charges $q_i$ are normalized such that the sum of all partial charges is an integer (i.e., the formal charge $q_{\text{formal}}$ of the molecule). This was achieved by subtracting the average predicted partial charge of a molecule from each partial charge, and adding the formal charge normalised to the number of atoms (see Eq. \ref{eq:gnn_normalize}). The latter term is necessary for formal charges $\neq 0$. 
A size of 200 was chosen for all hidden layers.
\begin{equation}\label{eq:gnn_normalize}
    q'_i = q_i +\frac{q_{\text{formal}}}{N_{\text{atoms}}} - \frac{1}{N_{\text{atoms}}} \sum_i^{N_{\text{atoms}}} q_i ,
\end{equation}
where $N_{\text{atoms}}$ is the number of atom (partial charges) in the molecule.

For the atom and bond embedding, an adapted version of the features proposed by Kearnes \textit{et al.} \cite{Kearnes2016_molecular_graph_convolution} was used. Atoms were encoded by creating a feature vector of length 23 containing element type (i.e., C, N, O, F, P, S, Cl, Br, I, or H), formal charge, hybridization (i.e., SP, SP2, or SP3), aromaticity, and degree (i.e., 0, 1, 2, 3, 4, 5, or other). Bonds were encoded by creating a feature vector of length 11 containing the bond type (i.e., single, double, triple, or aromatic), whether the bond is in a ring, whether the bond is conjugated, and stereo code following the RDKit \cite{Landrum2023_rdkit} definition (i.e., {STEREONONE}, {STEREOANY}, {STEREOE}, {STEREOZ}, or other). 

\subsubsection{Training Procedure}
\begin{figure}[H]
    \centering
    \includegraphics[width=0.99\textwidth]{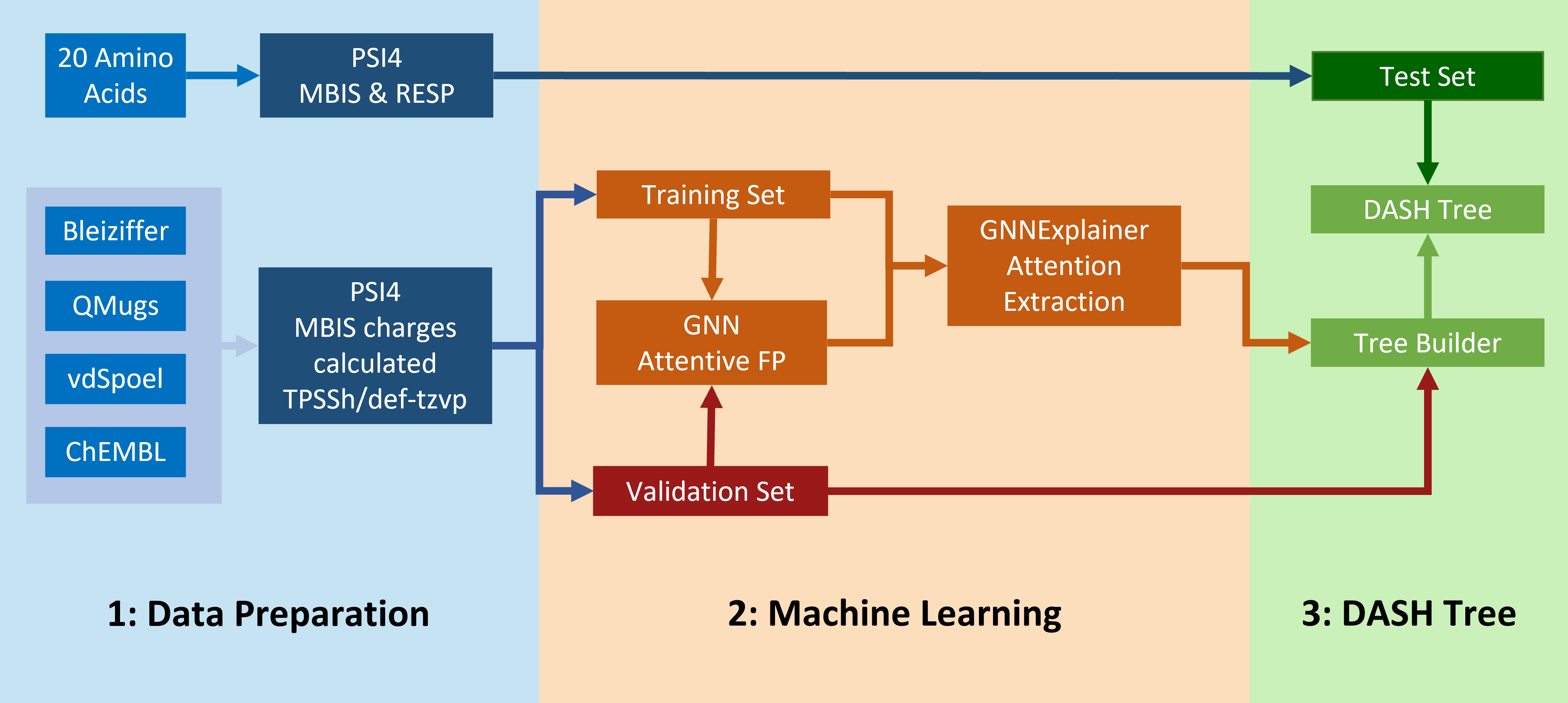}
    \caption{Schematic depiction of the workflow to construct the DASH tree structure: (1: Data Preparation) Reference charges were calculated from molecules from multiple data sets (blue). (2: Machine Learning) Molecules were split into a training set (orange) and a validation set (red). The training set was the input for training an Attentive FP GNN model to learn partial charges. The GNN was in turn the input for the GNNExplainer to extract attention values on the training set molecules. (3: DASH Tree) This data was subsequently used to construct the DASH tree structure. The GNN and DASH tree employed the same validation set. The test set (dark green) is only used for verification of the final DASH Tree.}
    \label{fig:workflow}
\end{figure}

The GNN was trained on all available conformers of a randomly selected 90\% subset (976081 3D structures) using the Adam optimiser \cite{Kingma2014_adam_optimizer} for 100 epochs. The mean squared error was chosen as the loss function. The effect of different learning rates (i.e., 0.0001, 0.00001, and 0.000001) and batch sizes (i.e., 32, 64, 128, 256, and 512) were studied in a hyperparameter optimization. 
The remaining 10\% of the data set (100'171 3D structures) were used as validation set during training of the GNN. The same split was later used for the DASH tree construction (see Figure \ref{fig:workflow}),

\subsection{Extraction of the Attention Values}
The attention values of the trained GNN were extracted with GNNExplainer from PyTorch Geometric \cite{Ying_GNNExplainer}. GNNExplainer takes as input the trained graph-based model, the data for which the attention should be extracted, and the number of epochs that the GNNExplainer model should be run on the data in order to generate attention values. The number of epochs was set to 500, the learning rate to 0.01, and the return type was set to the default value of {\tt{log\_prob}}. These values were found to give a good performance of the model, no systematic parameter search was performed. The attention values were then extracted for all atoms in all molecules in the training set.
Note that the attention values are not directly normalized per molecule. To enable the comparison of values between molecules, we divided the attention value of each atom in a molecule by the sum of all attention values in the molecule.

For a given atom, the neighbouring atoms contributing most to the prediction of its partial charge can be identified using either an attention threshold or a fixed number of atoms (environment size). The subgraphs (or substructure) of a molecule extracted in this manner can be compared to chemical intuition and can be processed further to generate a substructure-based table (i.e., using SMARTS or SMILES) to assign atomic partial charges of a molecule.
The choice of the metaparameter (attention threshold or the number of atoms) determines the performance of such an assignment table (accuracy versus speed of assignment).

\subsection{Dynamic Attention-based Substructure Hierarchy (DASH)}
To circumvent the issues associated with a static cutoff (either in the number of atoms or the attention), we propose a dynamic attention-based substructure hierarchy (DASH), where each node corresponds to a certain atom type, the neighbors of a node are ordered by attention, and branches can have different depths (dynamic). 
This way the attention values can be used to linearize the search through the exponentially growing number of possible patterns. To parameterize a particular atom in a molecule, a subgraph (substructure) of the molecule is grown starting from this atom by iteratively adding the neighboring atom with the highest attention value to the subgraph until a user-defined depth is reached. Note that no attention values have to be calculated for new molecule; the partial-charge assignment occurs by looking up the environments of each atom in the DASH.

\subsubsection{Atom Features}
To build the subgraphs, we need to define a feature vector for each atom that contains the necessary information to identify the atom type. While the same features as in the GNN training could be used, the feature vector for DASH should be as small as possible to reduce the number of possible patterns and improve human readability. We want to be able to calculate the atom features easily and rapidly from a RDKit molecule. Thus, we decided for an atomic feature vector with the following information:
\begin{itemize}
  \item Element type (H, C, N, O, S, F, Cl, Br, P, I, B)
  \item Number of bonds (1, 2, 3, 4, 5)
  \item Formal charge (-1, 0, 1)
  \item IsConjugated (True or False)
  \item Number of attached hydrogens (0, 1, 2, 3)
\end{itemize}
Only the isConjugated flag is set to true for an atom if at least one of its connecting bonds is conjugated.
This definition leads to 122 possible initial atom types, since many combinations of these properties are not physical or not present in the data set. For example, a hydrogen atom with one bond and a formal charge of zero has the atom type {\tt{``H 1 0 False 0''}}. These feature vectors are further translated to simple integers (keys) and stored in a dictionary object. 

\begin{figure}[H]
    \centering
    \includegraphics[width=0.99\textwidth]{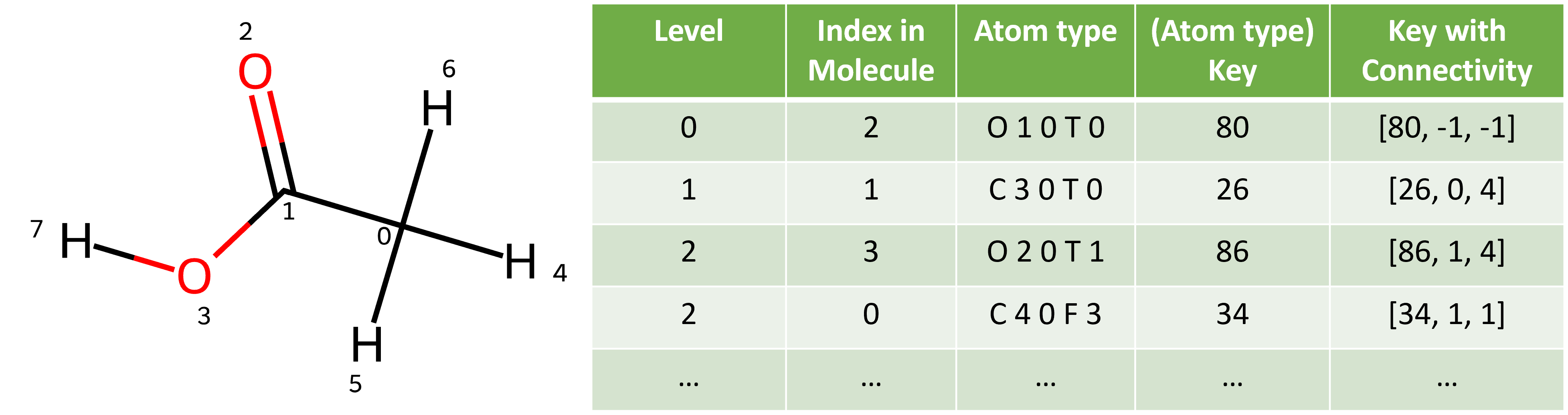}
    \caption{Example of the feature vectors / atom types of acetic acid. The oxygen atom with the index 2 is selected as root of the subgraph (level 0). Therefore, its connection information is non existing and set to -1. The second atom to be added is the carbon with index 1 and an atom type 26, which is connected to atom 0 in the subgraph with a conjugated double bond (type 4). Next, a choice has to be made between the oxygen with index 3 and the carbon with index 0 based on the attention values.}
    \label{fig:example_atom_feature}
\end{figure}

When the subgraph is extended by one atom during the DASH construction, the atom with the highest attention value is added. The feature vector for the new atom contains additional information about how it is attached to the current subgraph (i.e., relative index in the subgraph and bond type). The bond type is an integer with value 1 (single), 2 (double), 3 (triple), or 4 (conjugated, independent whether single or double bond). 
For the hydrogen dimer H$_2$ as example, the second hydrogen atom has the feature vector {\tt{[37, 0, 1]}} (atom-type key with connectivity), where the first element is the key of the atom type in the dictionary, the second element is the level of the atom in the subgraph (the root of the subgraph has level 0), and the third element is the bond type. The key with connectivity information can be used as an identifier for a node (atom), chained together to generate a chemical pattern.
For a more complex example, the atom feature vectors of acetic acid are shown in Figure \ref{fig:example_atom_feature}.

\subsubsection{DASH Implementation as Tree Structure}
Level 0 of the tree consists of the 122 nodes, one for each atom type, which branch out from the root. Every node stores the key of the atom type together with the GNN attention value and computed partial charge of every atom with that type in the training set. Level 0 could already be a simple look-up table for partial charges for a force field by simply averaging the partial charges at each node. This simple approach would, however, ignore most of the information about the environment of the atom and result in fairly crude partial charges. 

\begin{figure}[H]
    \centering
    \includegraphics[width=0.99\textwidth]{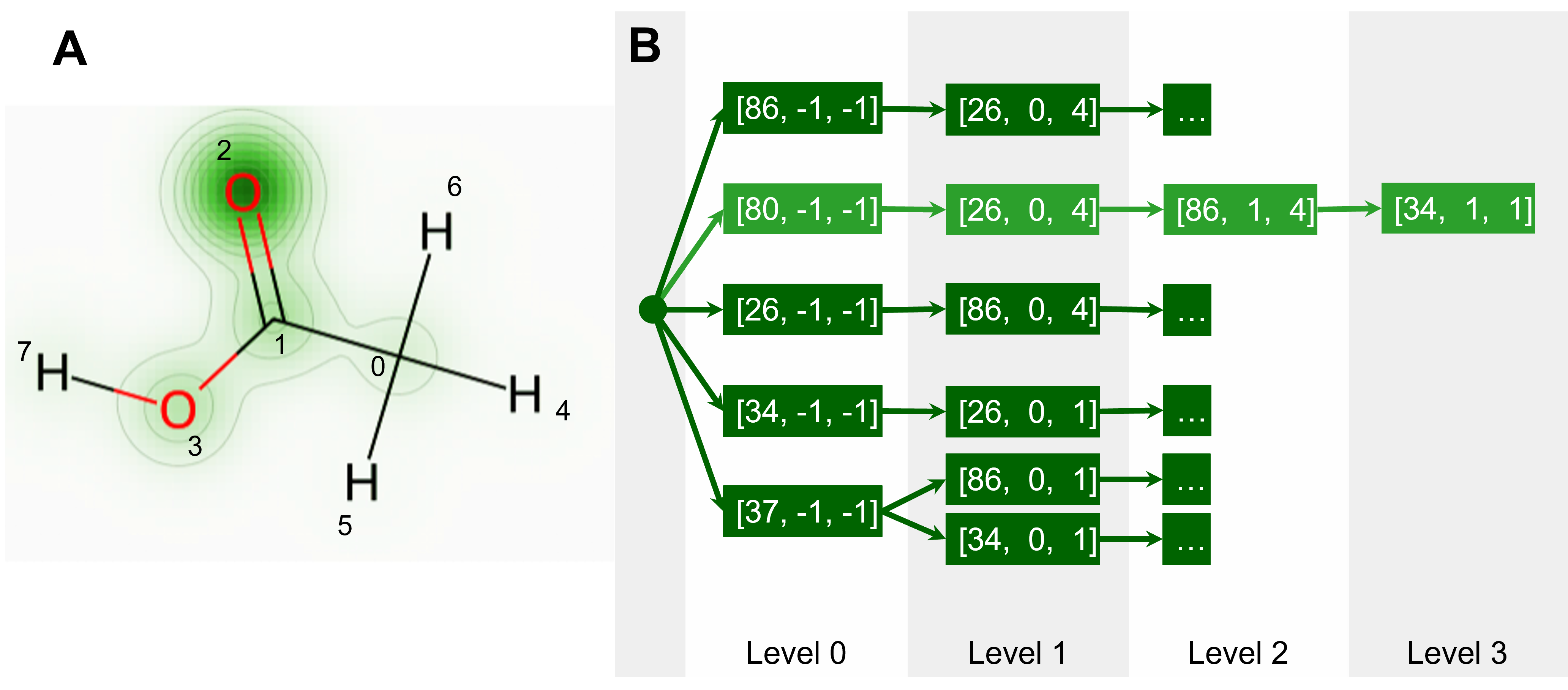}
    \caption{Example of acetic acid. (\textbf{A}): Attention values for the carbonyl oxygen atom (index 2, level 0) shown as a heatmap overlayed with the molecule. (\textbf{B}): DASH tree if it were constructed based only on acetic acid. In light green, the subgraph starting at the oxygen atom (index 2) is highlighted, with the atom types and connectivity information from Figure \ref{fig:example_atom_feature}. The remaining nodes of the DASH tree are shown in dark green. For clarity, nodes in higher levels are omitted, as indicated with ``...''. In this simple example, all heavy atoms could be uniquely identified on level 0. The branching of the hydrogens (atom type 37) into the two different subgraphs occurs in level 1.}
    \label{fig:example_tree_building}
\end{figure}

The accuracy can be improved by taking larger substructures into account, thus adding more information about the atomic environment. In the DASH approach, this is done by adding neighboring atoms in the order of decreasing attention values until the maximum graph depth is reached or all atoms in the training molecule have been added. 
Since each node in the tree stores not only a list of partial charges (from which an average charge per node can be calculated) but also the attention values, the attention could be used as an early stopping condition during the construction procedure to avoid overfitting of the tree to the training data set (not done here). 
A pseudocode implementation of the algorithm is provided in the Supporting Information. Figure \ref{fig:example_tree_building} shows an example DASH tree if it were constructed based on only one molecule (acetic acid). 
The final DASH tree was built with the same training data set as used for the GNN training to avoid any mixing of training and validation sets.

\subsubsection{Normalizing DASH Partial Charges}
As each atom is considered individually in the DASH assignment process, the resulting partial charges do not necessarily sum up exactly to the formal charge on the molecule. We explored two normalization schemes to address this issue. The first scheme calculates the difference between the sum of the partial charges and the formal charge on the molecule (Eq.~\ref{eq:tree_charge_difference}), divides this by the number of atoms, and adds the result to the partial charge of each atom (Eq.~\ref{eq:tree_normalize}). The second normalization scheme makes use of the standard deviation of the partial charges assigned to each atom. Here, the difference between the sum of the partial charges and the formal charge is distributed across the atoms using weights derived from the standard deviations of the partial charges assigned by DASH (Eq.~\ref{eq:tree_normalize_std}).
\begin{equation}
\label{eq:tree_charge_difference}
\Delta Q = \sum_{i=0}^{N} Q_i - Q_{\text{formal}}
\end{equation}
\begin{equation}
\label{eq:tree_normalize}
Q_i' = Q_i + \frac{\Delta Q}{N}
\end{equation}
\begin{equation}
\label{eq:tree_normalize_std}
Q_i' = Q_i + \frac{\Delta Q \cdot \sigma_i}{\sum_{j=0}^{N} \sigma_j}
\end{equation}
where $Q_i$ is the partial charge assigned to atom $i$ by the tree, $Q_{\text{formal}}$ is the formal charge on the molecule with $N$ atoms, and $\sigma_i$ is the standard deviation of the partial charges in the leaf of the DASH tree corresponding to atom $i$ from the tree. Both methods were tested on the validation set and compared to both the QM reference charges and the raw DASH partial charges.

\subsubsection{Symmetrizing DASH Partial Charges}
The QM reference charges are, per definition, dependent on the 3D conformation of the molecule. This means that topologically equivalent atoms can have different computed partial charges. This conformational dependency is in principle removed in the GNN due to the 2D input (topology), i.e., partial charges of topologically equivalent atoms will be averaged in the GNN predictions. However, asymmetries may be (re)introduced in the DASH assignment process because  subgraphs are matched using a greedy approach (always adding the node with the highest attention). 
The degree of asymmetry can be tested using the RDKit {\tt{CanonicalRankAtoms}} function to find atoms with the same rank (topologically equivalent), 
and compare the partial charges from the QM reference calculation, the GNN, and DASH. Note that the DASH partial charges can be simply symmetrized by averaging the partial charges of the atoms with the same rank.

\subsubsection{Assigning DASH Partial Charges for New Molecules}
\begin{figure}[H]
    \centering
    \includegraphics[width=\textwidth]{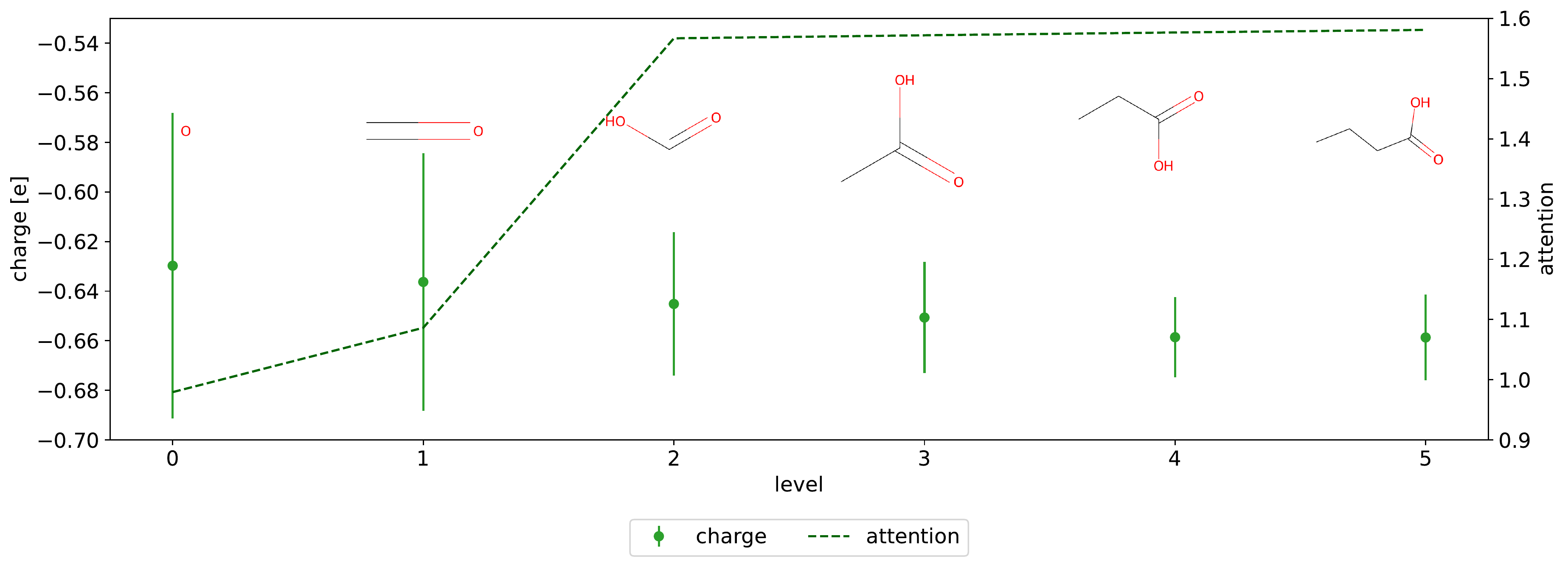}
    \caption{Example of the DASH charge assignment for matching "=O" in the molecule CCCC(=O)O. By traversing the levels of the DASH tree, a fragment for the charge assignment is built up. The partial charge at each level is denoted by a green dot with standard deviation (left $y$-axis). The attention value is shown with a green dashed line (right $y$-axis). 
    }
    \label{fig:example_tree_matching}
\end{figure}

DASH partial charges of a new molecule are assigned by first matching each atom separately in the DASH tree structure. For each atom, the tree is traversed until either the maximal depth or the attention threshold is reached, or the subgraph is equal to the size of the molecule. In Figure \ref{fig:example_tree_matching}, the assignment process is shown for the double-bonded oxygen in butyric acid as an example. The subgraph is built up over six levels, where the nodes at each level contain partial charges with standard deviations and attention values (which are used as the stopping criterion).
After all atoms have partial charges assigned individually, the atomic charges are normalized and symmetrized.

\subsubsection{OpenFF Plug-in}
To integrate the DASH partial charges with the rest of the force-field assignment, the DASH tree structure was implemented as a {\tt{NonBondedHandler}} in the OpenFF toolkit software \cite{Mobley2018EscapingPerception}, and can be installed as a plug-in. The plug-in, the stand-alone DASH and the DASH tree constructor functions as well as the GNN are available as open source source code on GitHub (\url{https://github.com/rinikerlab/DASH-tree}).

\subsection{Performance Assessment}
The prediction accuracy with DASH was assessed using the same validation set as for the GNN as well as an external test set consisting of the 20 canonical amino acids. The latter data set shows the potential applicability of DASH for biomolecular force fields. Two different meta-parameters were compared: the maximal depth of the DASH tree structure, and the attention threshold when constructing DASH. In addition, simple pruning by the maximal depth or attention threshold was compared to a pruning scheme based on the standard deviation of the partial charges in the nodes. If the change in the average partial charge from parent to child was smaller than the standard deviation of all partial charges in the child nodes multiplied by a scaling factor, the node was pruned (i.e., $\left(\frac{1}{{N_{\text{parent}}}} \sum_i^{N_{\text{parent}}} q_i - \frac{1}{{N_{\text{child}}}}\sum_j^{N_{child}} q_j \right) < s \cdot \sigma(q_i^{\text{parent}})$). Different scaling factors were tested.

The performance was assessed using the mean absolute error (MAE), the root-mean-squared error (RMSE), and the Pearson correlation coefficient $R^2$ compared to the QM reference partial charges. In addition, the computing time needed to assign the DASH charges and the size of the DASH tree itself was monitored.

\subsection{Other Partial-Charge Models}
The DASH partial charges were compared with semi-empirical Mulliken-type charges \cite{Mulliken}, AM1-BCC charges \cite{AM1BCC}, 2D Gasteiger charges \cite{Gasteiger19783181_atomiCharge}, and MMFF94 partial charges \cite{MMFF94}. The Mulliken-type charges were taken from the XTB-GFN2 \cite{XTBGFN2} conformer optimization step during data preparation. AM1-BCC charges were calculated with the OpenFF toolkit (version 0.10.0) \cite{Mobley2018EscapingPerception}, using the Amber toolkit (version 22.0) \cite{Case2022AMBER2022}. The 2D Gasteiger and MMFF94 partial charges were obtained with RDKit \cite{Landrum2023_rdkit} (version 2022.9.1).
For the amino acid test set, RESP charges \cite{RESP} calculated with PSI4 and PsiRESP (b3lyp/sto-3g, RESP2) were also compared.

\subsection{Liquid Properties: MD Simulations}
MD simulations were performed for a set of 123 organic liquids with experimental values for density and heat of vaporization available. The molecules were parameterized using the OpenFF toolkit \cite{Mobley2018EscapingPerception} with OpenFF version 2.0.0 (Parsley) \cite{openff_parsley} and the DASH plug-in. To evaluate the density and heat of vaporization, the openFF-evaluator \cite{boothroyd2022open_evaluator} package was used, with the default schemes to estimate the two properties in the OpenMM \cite{Eastman2017_openmm} engine (version 8.0.0). The default scheme uses a box of 1000 molecules and consists of an energy minimization, a NPT equilibration (100'000 steps with 2~fs), and up to 100 NPT production runs (1'000'000 steps with 2~fs) until a convergence criteria is met, followed by a de-correlation step. All simulations in this scheme were performed with the Langevin integrator (298.15~K) and a Monte Carlo barostat (101.325~kPa) for the NPT simulations. The results were compared to a simulation using AM1-BCC charges.

\section{Results and Discussion}

\subsection{Overview of the Data Set}
\begin{figure}[H]
  \centering
  \includegraphics[width=0.7\textwidth]{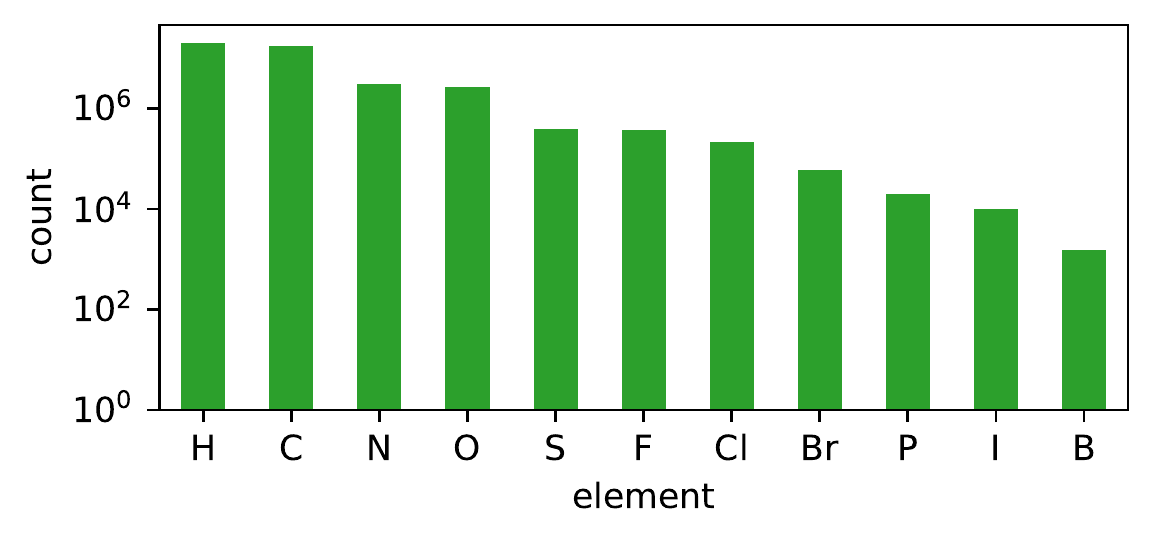}
  \caption{Atom counts per element in the full data set (393'692 unique molecules).}
  \label{fig:data set_distribution_mf}
\end{figure}

The final data set contained 393'692 unique molecules with up to three conformers per molecule. These molecules were selected to represent the substructures (as measured by unique bits in Morgen fingerprint of radius 2 (MFP2)) found in molecules with a maximum molecular weight of 500 g/mol in ChEMBL with a minimal subset of molecules. Figure \ref{fig:data set_distribution_mf} shows the number of data points per element. While the goal was that each bit is represented at least five times in the data set, a few MFP2 bits are only present once. These belong to very small molecules for which radius 2 describes the entire molecule (i.e., there exists exactly one molecule that can have this bit). 

\begin{figure}[H]
  \centering
  \includegraphics[width=0.9\textwidth]{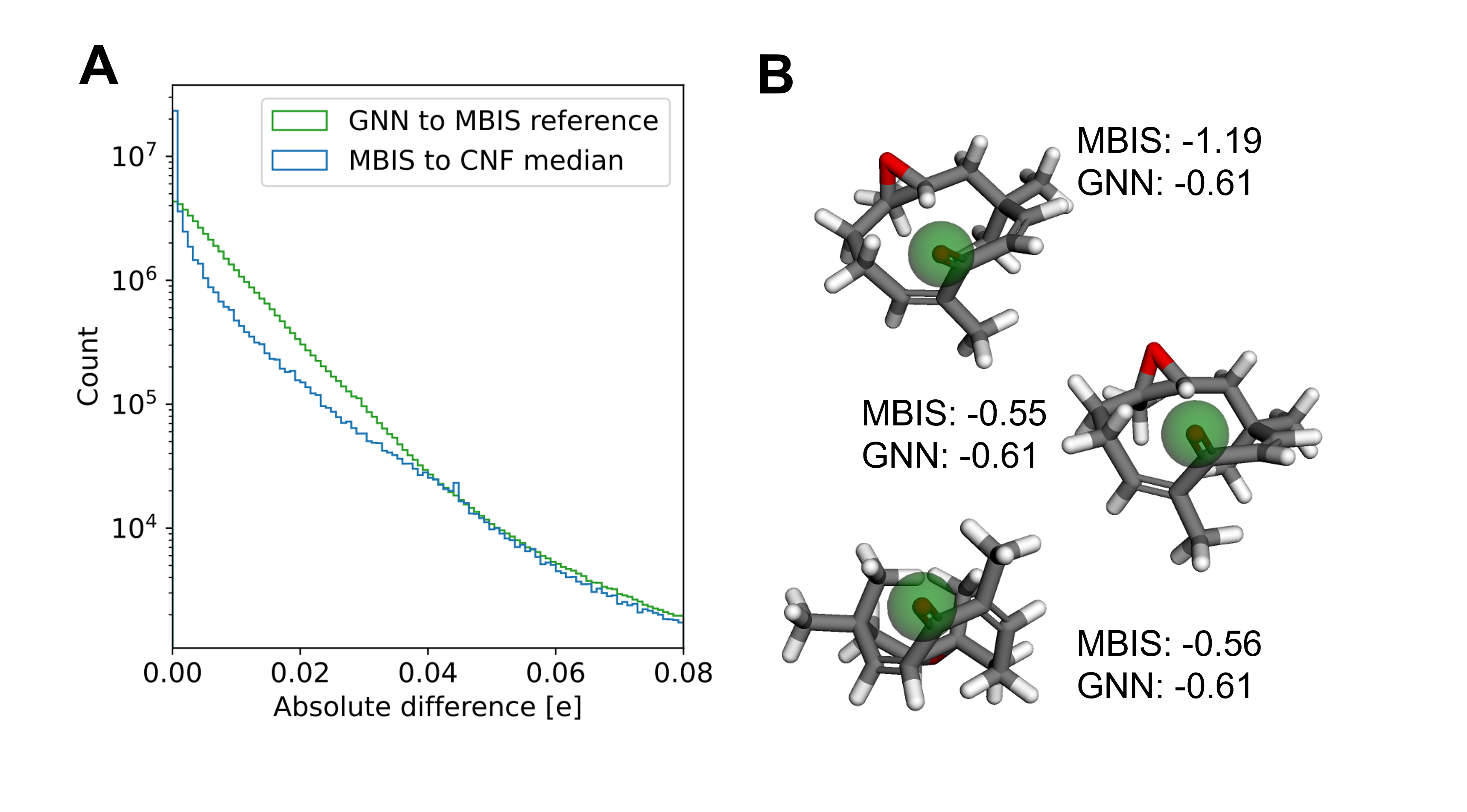}
  \caption{(\textbf{A}): Histogram of absolute partial-charge differences for all molecules in the full set. MBIS reference (blue): For each atom, the difference between each conformer and the median of the three conformers (CNF) is shown. GNN (green): For each atom, the difference of the GNN prediction to the individual MBIS reference. (\textbf{B}): Example molecule from the data set where the three conformers have very different MBIS partial charges. The atom with the largest differences is marked with a green sphere.}
  \label{fig:charge_diff_hist}
\end{figure}

To estimate the conformational variation of the partial charges, we compared the differences in the MBIS reference charges between the three conformers of the same molecule. The distribution of the absolute differences is shown in Figure \ref{fig:charge_diff_hist}. The RMSE of the individual conformer to the median over the three conformers is $0.0125\,e$, which presents an upper bound on the accuracy that can be reached by a ML model which uses the 2D topology of the molecule as input. In this case, the RMSE of the GNN with respect to the individual MBIS reference charges is $0.014~e$ over the full data set. Figure \ref{fig:charge_diff_hist}B shows an example molecule where the three conformers have very different MBIS partial charges due to large differences in the conformations. One of the atoms with the largest absolute difference is the oxygen atom marked with a green sphere, which is in close proximity to very different functional groups in each conformer. 

The data set of 393'692 unique molecules (three conformers each, i.e., 1'076'252 3D structures in total) was split randomly into a 90\% subset for training of the GNN and DASH, while the remaining 10\% (100'171 3D structures) served as validation set.

\subsection{GNN Performance}

\begin{figure}[H]
  \centering
  \includegraphics[width=\textwidth]{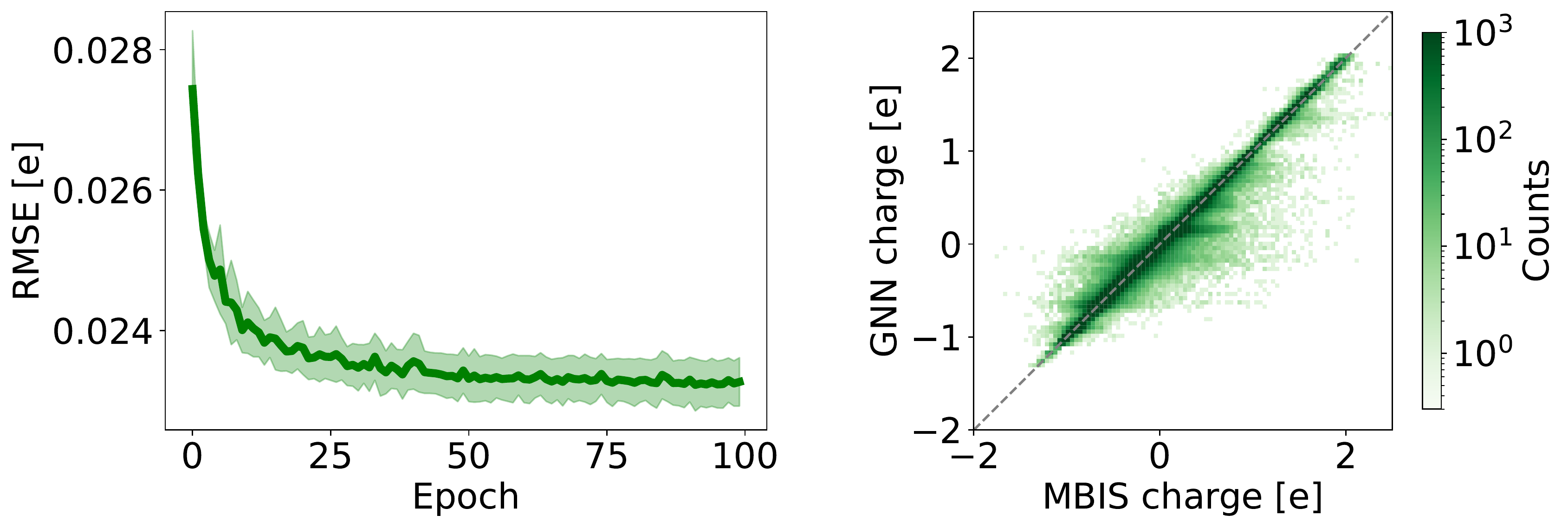}
  \caption{(Left): RMSE of the GNN predicted charges with respect to the MBIS reference charges on the validation set (100'171 3D structures) as a function of the training epoch. The GNN was trained with a learning rate of 0.0001 and a batch size of 64 (final RMSE~=~$0.0238\pm0.0004$\,e). The shaded area indicates the statistical uncertainty. (Right): Comparison between the GNN predicted charges (100 epochs) and the MBIS reference charges on the validation set.}
  \label{fig:Bestlossdecay}
\end{figure}

The architecture of the GNN was already optimized in Ref.~\cite{Xiong2020_AttentiveFP}, i.e., it was kept constant in this study. The training parameters (learning rate from 0.000001 to 0.01 with 10-fold increments and batch size ranging from 64 to 512 with 2-fold increments) were screened to identify optimal values for the data set (Table S1 in the Supporting Information). 
A learning rate of 0.0001 and a batch size of 64 yielded the GNN model with the smallest RMSE on the validation set (left panel of Figure \ref{fig:Bestlossdecay}).
The distribution of the absolute differences between the GNN prediction and the MBIS reference is shown in Figure~\ref{fig:charge_diff_hist}A, reaching an accuracy similar to the conformational variation limit. The direct comparison is provided in the right panel of Figure~\ref{fig:Bestlossdecay}.

\subsection{DASH Performance}
\begin{figure}[H]
  \centering
  \includegraphics[width=\textwidth]{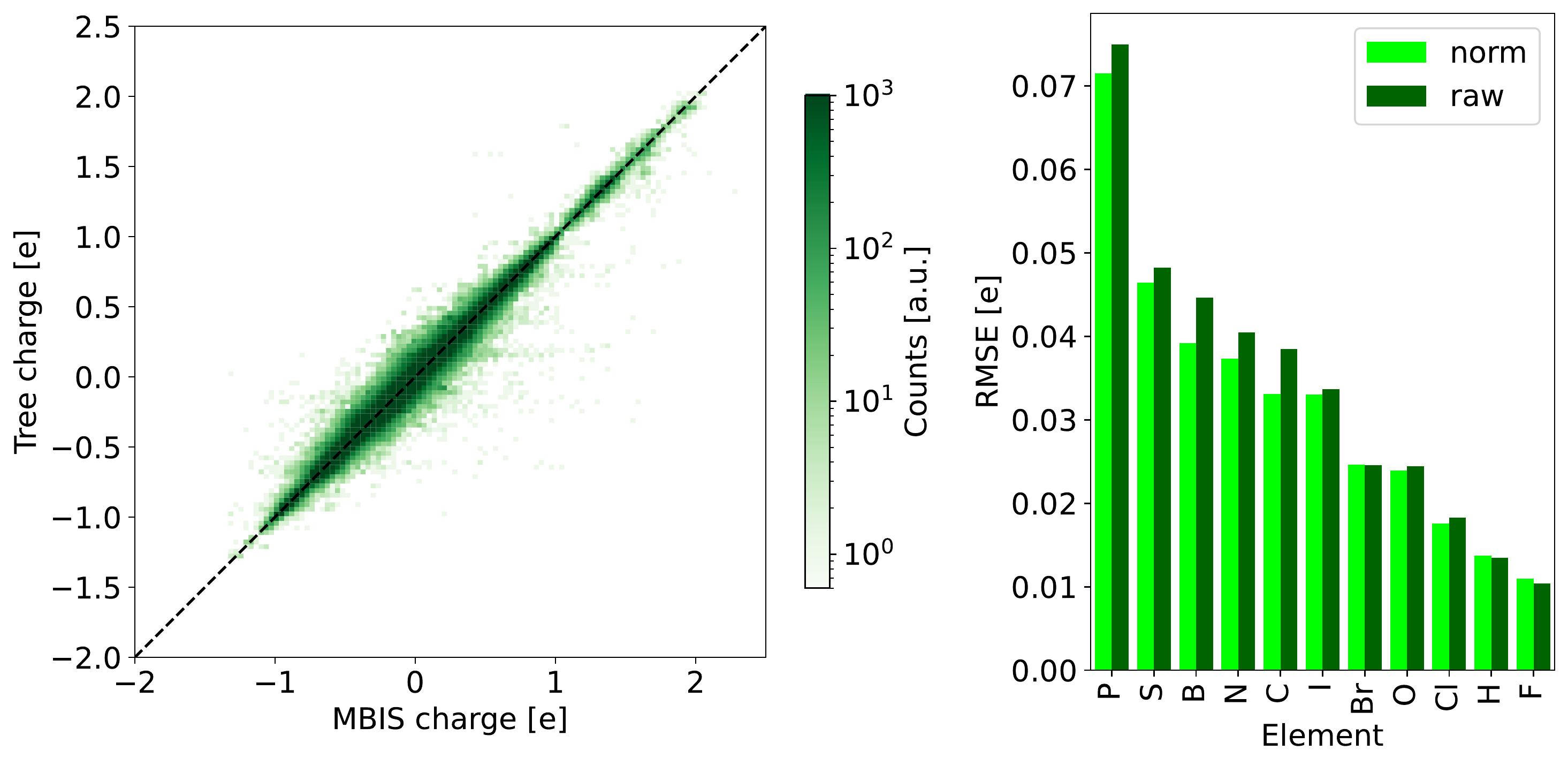}
  \caption{(Left): Comparison between the normalized DASH partial charges and the MBIS reference charges on the validation set (100'171 3D structures). A maximal depth of 16 layers and an attention threshold of 5.23 was used to construct the DASH tree structure. The color of the points indicates the number of atoms in a pixel. (Right): RMSE of the DASH partial charges with respect to the MBIS reference charges for each element on the validation set. The 'raw' DASH charges are shown in dark green and the normalized ones (Eq.~\ref{eq:tree_normalize_std}) in light green.}
  \label{fig:tree_test152_SingleTreeCorrelation}
\end{figure}

After the GNN was successfully trained and tested, the attention values were extracted and the DASH tree structure was constructed using the training set of the GNN. The validation set was used to evaluate the performance of DASH and to tune its hyperparameters: maximal depth and attention threshold. A tree with a maximal depth of 16 layers and an attention threshold of 0.95 was found to perform well on the validation set (Figure \ref{fig:tree_test152_SingleTreeCorrelation}). The RMSE as a function of the maximal depth is provided in Figure S1 in the Supporting Information. 
The same figure also shows that the time to assign partial charges increases roughly linear with the maximal depth. The choice of this hyperparameter is thus a trade-off between accuracy and speed of assignment.

The right panel of Figure \ref{fig:tree_test152_SingleTreeCorrelation} shows the RMSE of the DASH partial charges with respect to the MBIS reference charges on the validation set for each element. The RMSE values are generally very small. The largest RMSE values are observed for phosphorous, which is particularly difficult for charge assignment because it shows a large range of partial charges and it is underrepresented in the data set (and generally in ChEMBL).
Normalizing the DASH charges with Eq.~\ref{eq:tree_normalize_std} reduces the errors slightly. Using Eq.~\ref{eq:tree_normalize} instead gives very similar results (Figure S2 in the Supporting Information). 
As integer values for the total charge of a molecule are important for MD simulations, we used the normalization with Eq.~\ref{eq:tree_normalize_std} in the remainder of this work.

\begin{figure}[H]
  \centering
  \includegraphics[width=0.99\textwidth]{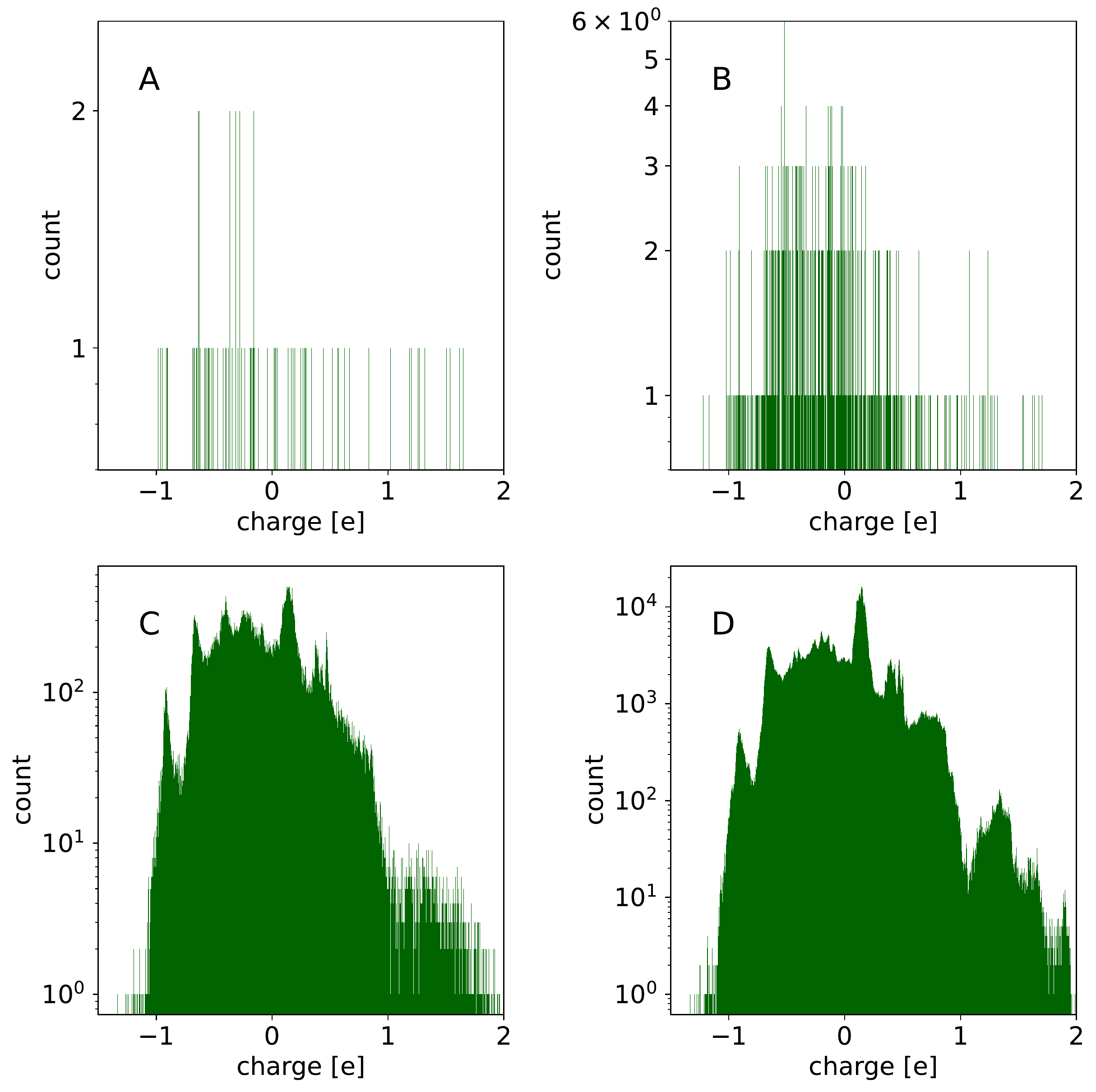}
  \caption{Distribution of the nodes in the DASH tree structure over the range of partial charges for tree level = 0 (A), 1 (B), 8 (C), and 16 (D). A bin size of 1/2000~e was used.}
  \label{fig:tree_branch_2dhist_double_plot}
\end{figure}

The effect of the depth of the DASH tree structure can also be seen in Figure \ref{fig:tree_branch_2dhist_double_plot}, which shows the distribution of the nodes over the range of partial charges at different levels of the DASH tree. Note that level 0 consists of the 122 initial atom types, which can be seen as discrete bars in the histogram. At level 1, the possible partial charges have already a much large set of possible, but still clearly separated values. This strong discretization becomes further refined with increasing depth. At the maximum depth of 16, there is more continuous distribution of the $4\cdot10^6$ nodes over the full range of possible partial charges. 

\begin{figure}[H]
    \centering
    \includegraphics[width=1.0\textwidth]{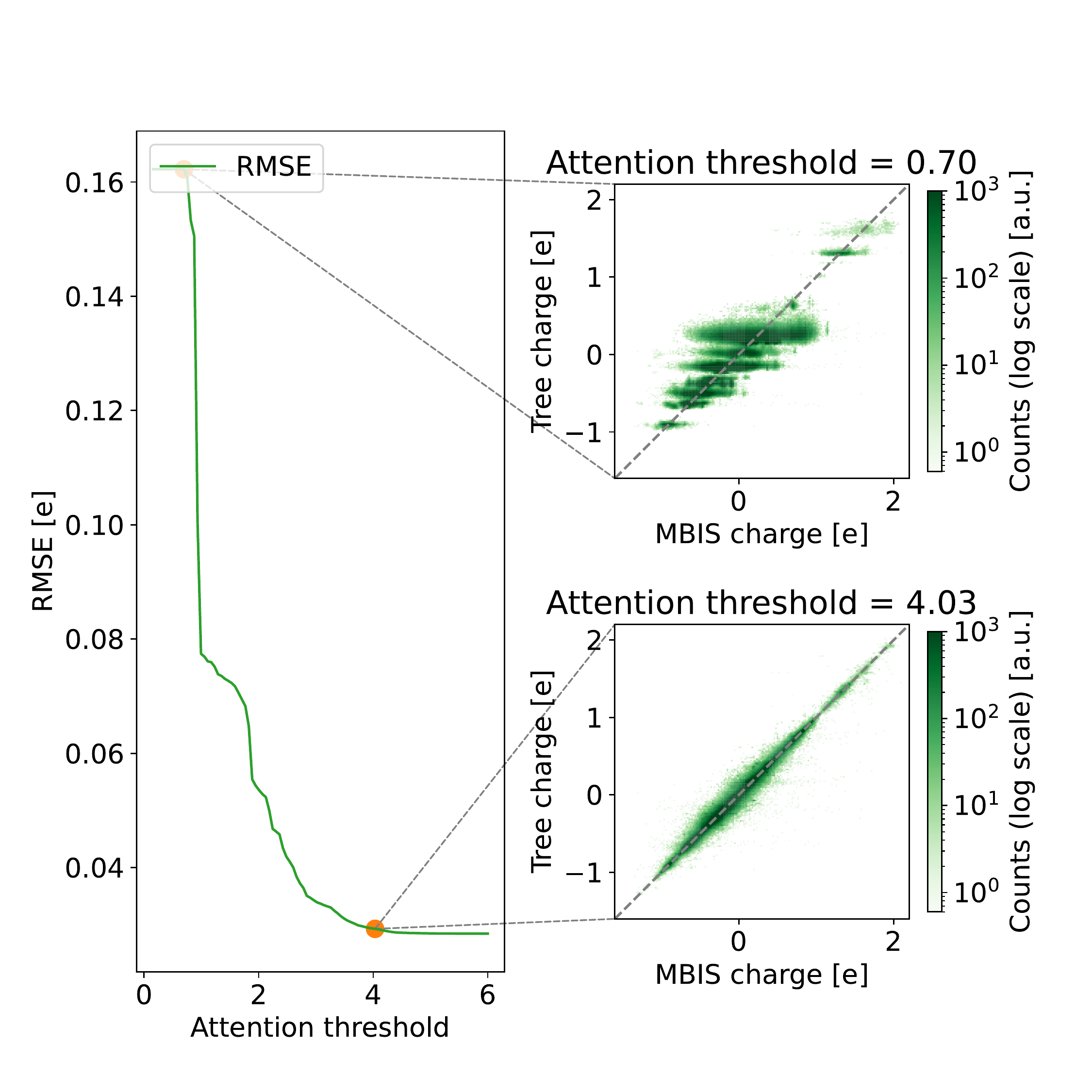}
    \caption{RMSE of the DASH partial charges with respect to the MBIS reference charges on the validation set as a function of the attention threshold. The minimum RMSE is at an attention threshold of 5.2. For two attention thresholds (0.7 and 4.03), the comparison between the DASH partial charges and the MBIS reference charges is shown on the right. A maximal depth of 16 was used for the DASH tree structure.}
    \label{fig:tree_AttentionVsRMSE_test_mols}
\end{figure}

A similar trend from a more discrete to a more continuous prediction of the partial charge can be seen in Figure \ref{fig:tree_AttentionVsRMSE_test_mols} as a function of the attention threshold. Note that the attention in the nodes is not strictly confined to the range of 0 to 1 because nodes can contain information from multiple molecules and the attention is normalized in the GNN over all atoms in a molecule. The RMSE of the DASH partial charges with respect to the MBIS reference charges on the validation set decreases with increasing attention threshold, reaching a minimum at around 5.2 (the exact value will depend on the training set). Interestingly, the RMSE converges to a value slightly above the minimum when increasing the attention threshold further. The fact that additional information does not improve the accuracy anymore is an indication that the tree is already able to capture all relevant information at this point and that the additional information leads to overfitting. 

\begin{figure}[H]
    \centering
    \includegraphics[width=0.5\textwidth]{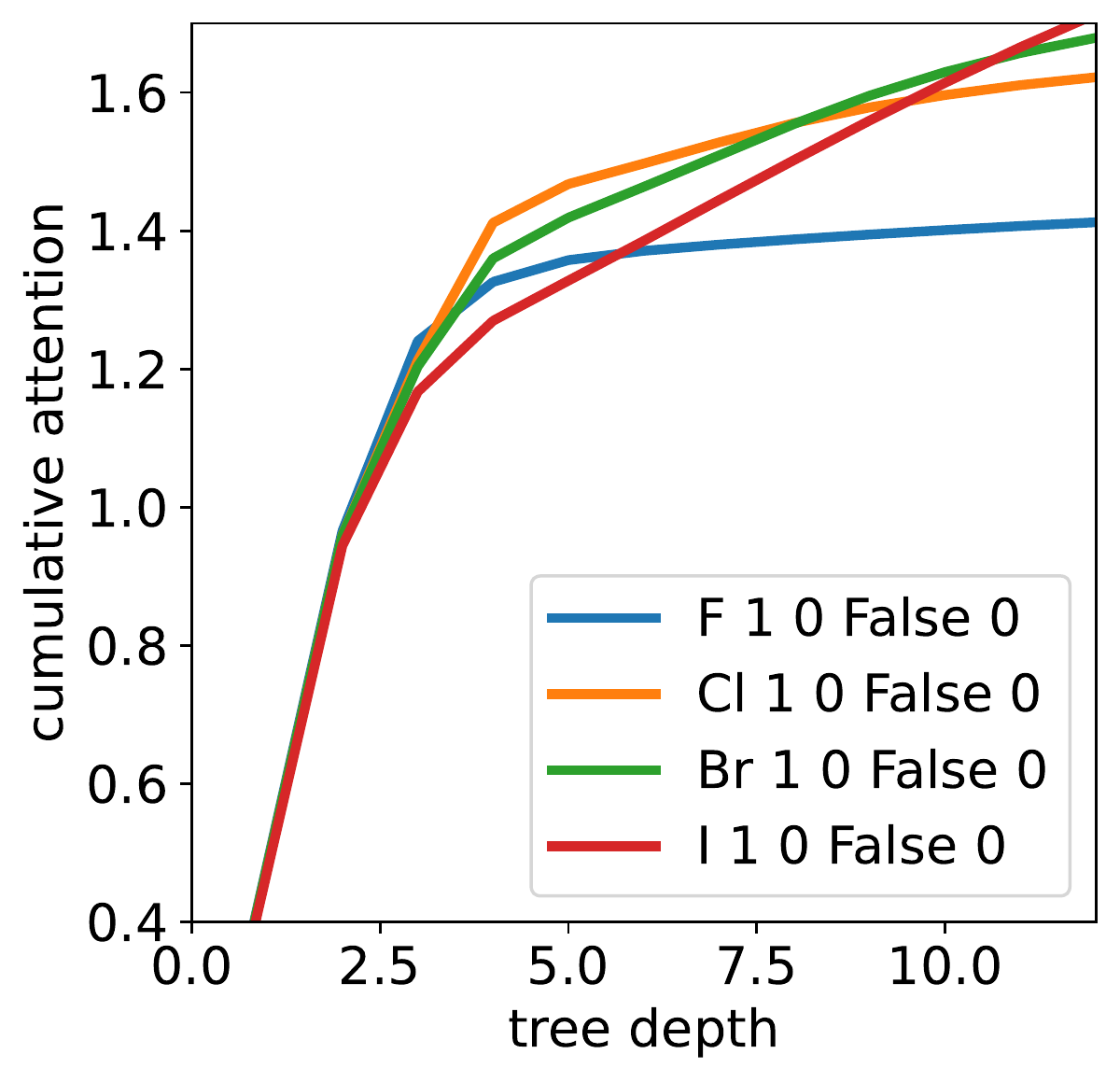}
    \caption{Cumulative attention of the nodes as a function of the DASH tree depth for the four halogen atom types.}
\label{fig:cumulativeAttentionPerLevelForHalogenBranches}
\end{figure}

The initial attention values of the 122 atom types (nodes at level 0 in the DASH tree structure) are shown in Figure S3 in the Supporting Information. 
Our assumption is that atom types with a high initial attention need little additional information from the environment for a good prediction of the partial charge, while atom types with a low attention values require larger subgraphs for a precise prediction.
It is possible to observe some chemical trends by comparing selected atom types. For example, the attention of the four atom types for the halogens show that fluorine atoms have the steepest increase in attention with increasing depth of the DASH tree, while iodine atoms have the slowest initial increase in attention (Figure \ref{fig:cumulativeAttentionPerLevelForHalogenBranches}). 
This slower increase in attention may be explained by the lower hardness of iodine compared to fluorine, and therefore the stronger influence of the environment on the partial charge.

When assigning the partial charge of an atom with the DASH tree structure, a greedy approach is followed, i.e., at each step the neighboring atom with the highest attention value is added to the subgraph. This can mean that topologically symmetric atoms may have different partial charges assigned. However, such asymmetries are rare and small (Figure S4 in the Supporting Information). 
To resolve this issue, a symmetrization step (i.e., averaging the partial charges of the symmetric atoms) was added after the normalization. This does not decrease the RMSE significantly (reduced by 0.12\%) on the validation set.

\subsection{Comparison with Other Partial-Charge Models}
\begin{figure}[H]
  \centering
  \includegraphics[width=1.0\textwidth]{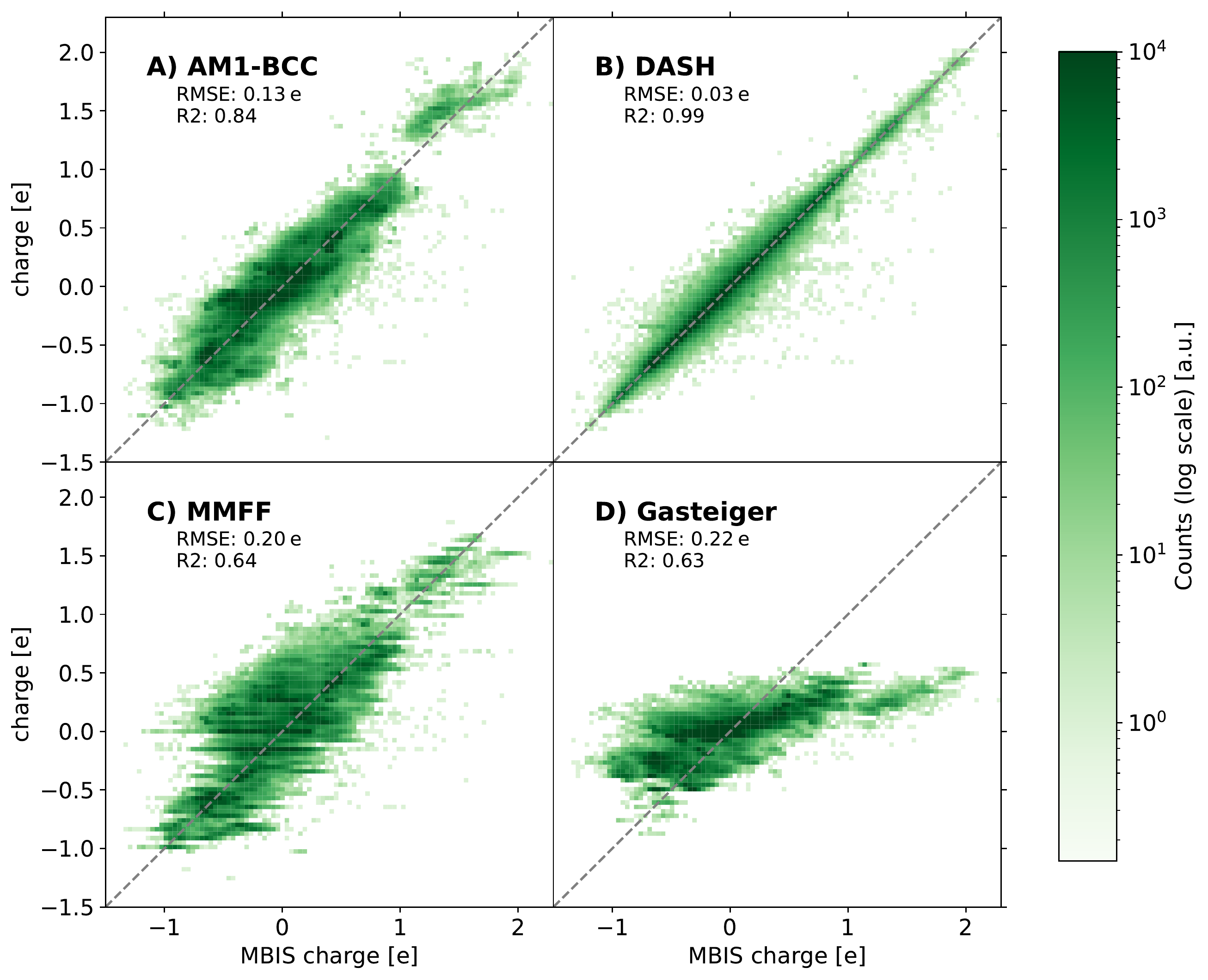}
  \caption{Comparison between the estimated partial charges with AM1-BCC (\textbf{A}), DASH (\textbf{B}), MMFF94 (\textbf{C}), and Gasteiger (\textbf{D}) and the MBIS reference charges on the validation set. The corresponding RMSE and R$^2$ values are shown in the subplots.}
  \label{fig:4charges_validationSet}
\end{figure}

First, we compared the accuracy of the DASH partial charges on the validation set with the performance of AM1-BCC, Gasteiger, and MMFF94 methods (Figure \ref{fig:4charges_validationSet}).  
Unsurprisingly, the Gasteiger partial charges have the lowest accuracy and are clearly not suited for MD simulations of condensed-phase systems. The weaker polarization results in a narrower range of partial charges (approximately between -0.5~e and +0.5~e). While AM1-BCC and MMFF94 charges are reasonably close to the MBIS reference charges. Both methods show larger deviations than the DASH partial charges (part of this may be because they were fitted to other reference charges), especially for slightly charged carbons in large conjugated systems where the partial charge is influenced by far away atoms. Interestingly, some discretization effects can be observed for the MMFF94 charges (visible as horizontal lines in the figure) due to the limited number of atom types in MMFF94.

\begin{figure}[H]
  \centering
  \includegraphics[width=1.0\textwidth]{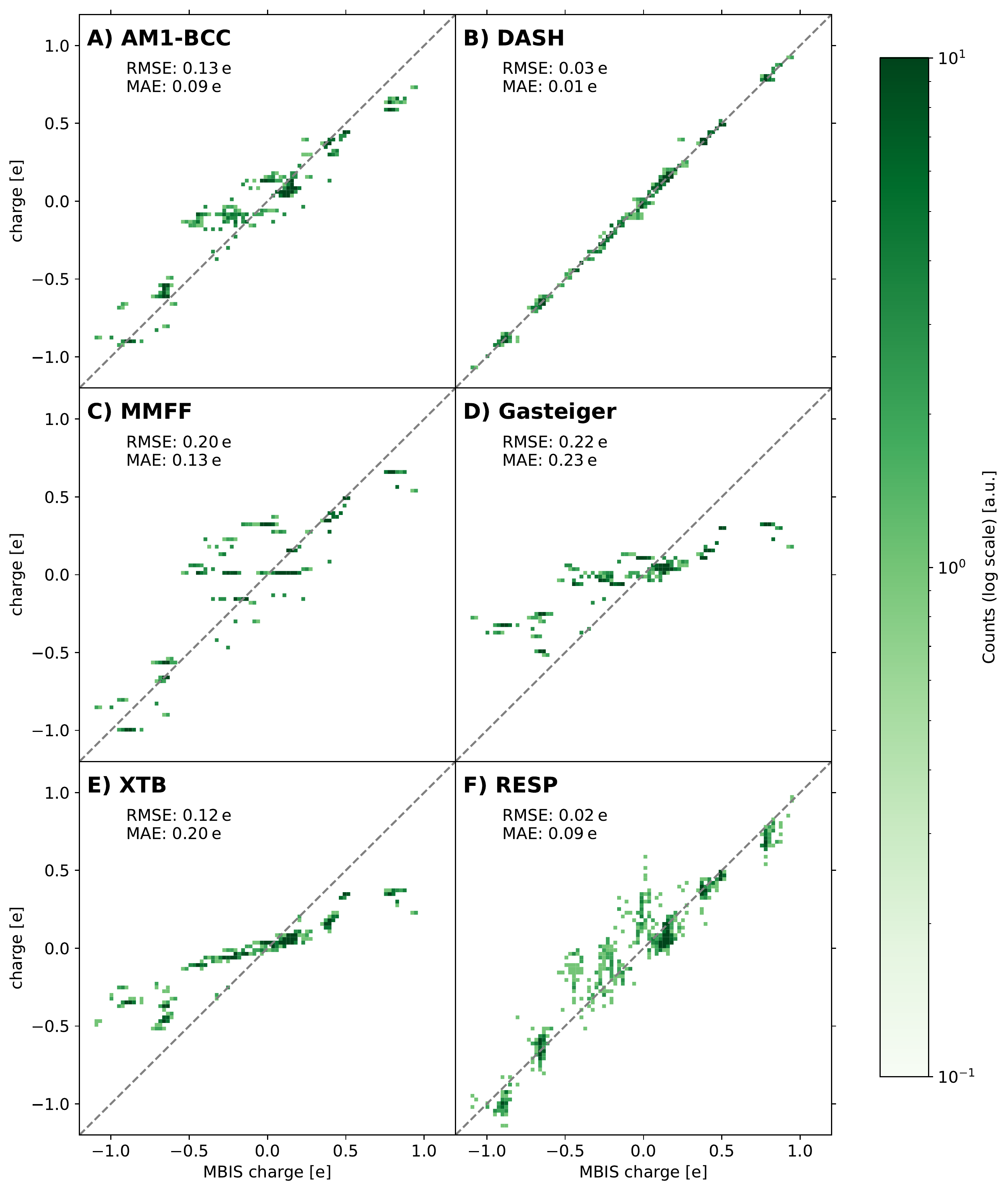}
  \caption{Comparison between the estimated partial charges with AM1-BCC (\textbf{A}), DASH (\textbf{B}), MMFF94 (\textbf{C}), Gasteiger (\textbf{D}), Mulliken-type from XTB-GNF2 (\textbf{E}), and RESP (\textbf{F}) and the MBIS reference charges on the external test set with the 20 canonical amino acids. Note that a different functional is used for the standard RESP partial charges than for the MBIS charges.}
  \label{fig:4charges_aaSet}
\end{figure}

Next, we compared the different partial-charge models for an external test set that consists of the 20 canonical amino acids. The motivation behind this data set is the usage of the DASH partial charges in protein simulations in the future. For this smaller data set, we calculated also RESP charges and Mulliken-type charge from the XTB-GFN2 optimization. The results are shown in Figure \ref{fig:4charges_aaSet}. The Mulliken-type charges from the XTB-GFN2 optimization show a similarly narrow range as the Gasteiger charges due to smaller polarization, which indicates that they are also not suited for fixed charge MD simulations of condensed phase systems.
RESP charges show overall a similar behaviour to the MBIS charges, although deviations can be observed for some atoms. Reasons for this could be the different functional and basis set typically used for RESP (B3LYP/STO-3G) compared to the MBIS charges extracted in this work and/or the stronger conformational dependency of RESP compared to MBIS  \cite{Verstraelen2016_mbis}.

\subsection{Liquid Properties}
\begin{figure}[H]
  \centering
  \includegraphics[width=1.0\textwidth]{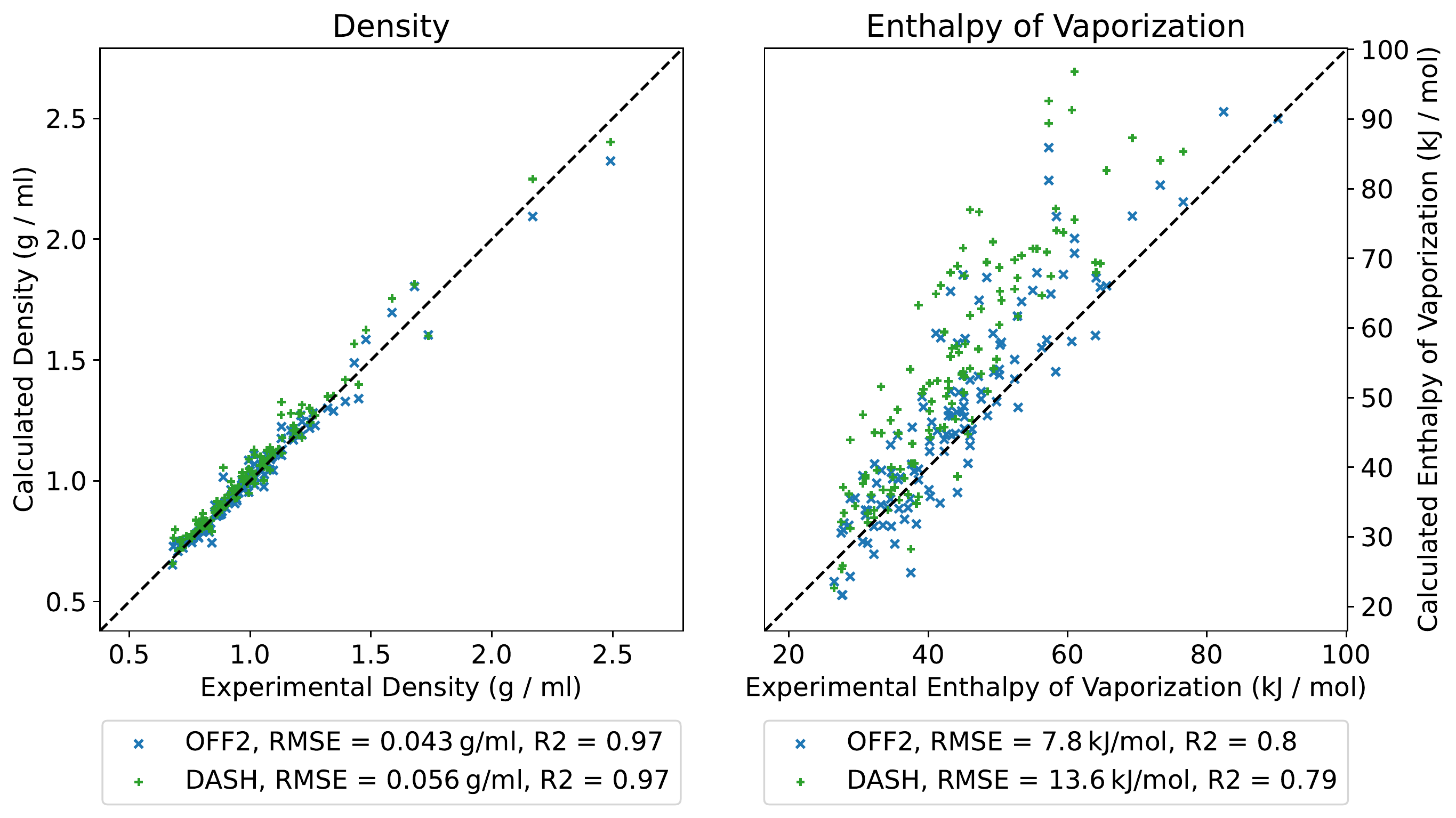}
  \caption{Comparison of the experimental density (left) and heat of vaporization (right) values of 123 organic liquids with the calculated values using OpenFF-2.0.0 with the default AM1-BCC charges (blue) and the DASH partial charges (green).}
  \label{fig:densHvap_treeAndOff}
\end{figure}

Finally, we tested the combination of DASH partial charges with the OpenFF-2.0.0 force field by calculating liquid properties (density and heat of vaporization) of 123 organic liquids with experimental data available \cite{caleman2012vanDerSpoel_data}. To construct the topologies, the DASH plug-in was used in the OpenFF workflow.
The comparison between the calculated values (using either DASH or AM1-BCC charges) and the experimental properties are shown in Figure \ref{fig:densHvap_treeAndOff}. Given that the Lennard-Jones parameters were not intended for DASH partial charges (i.e., no re-fitting was performed), the performance is similar with only a small increase in the RMSE values. For the heat of vaporization, there seems to be a slight shift towards larger values (overestimation). Nevertheless, the overall good agreement indicates that the DASH partial charges can be combined with the OpenFF-2.0.0 force field for condensed-phase simulations, with a substantially reduced computing time for the charge assignment (see below).

\subsection{Timings}

In addition to a high accuracy with respect to the MBIS reference, the DASH  partial charges can be assigned much faster (Table \ref{tab:computation_time}) than with commonly used methods such as AM1-BCC, and even four orders of magnitude faster than MBIS. While the assignment with Gasteiger and MMFF94 is even faster than DASH, the resulting partial charges are not well suited for MD simulations (as discussed above). Note that the assignment with DASH was carried out as a sequential single-thread program. The matching of different atoms in a molecule could in principle be done in parallel, potentially decreasing the computation time.

The required storage space to save all required data is slightly larger for DASH (about $\approx500MB$ if saved as CSV files per branch and zipped, or $\approx150MB$ stored as compressed pickle files for each branch) compared to the Pytorch state-dict of the GNN (7MB). However, on most modern machines with TBs of storage, these differences should not affect any performance. Note that the state-dict is a compressed machine format, while the CSV is a human readable file and contains extra information, which is redundant to machines. Furthermore, the GNN requires Pytorch and Pytorch-Geometric libraries, while DASH relies only on RDKit to interpret the molecules and assign the simple feature vectors required by the tree. This basic RDKit functionality is reasonably stable between versions of the toolkit, so we anticipate that it should not be necessary to regenerate the DASH model for each new RDKit release.

\begin{table}[H]
    \centering
    \begin{tabular}{c | c}
    \hline
       \textbf{Method} &  \textbf{Time [s]} \\ \hline \hline
         MMFF &  $1.04\cdot10^{-02}$ \\
    Gasteiger &  $2.45\cdot10^{-02}$ \\
          GNN &  $1.02\cdot10^{+00}$ \\
         DASH &  $3.87\cdot10^{+00}$ \\
      AM1-BCC &  $1.95\cdot10^{+02}$ \\
         MBIS &  $8.49\cdot10^{+03}$ \\
         RESP &  $1.10\cdot10^{+04}$ \\ \hline
    \end{tabular}
    \caption{Computation time required for different charge assignment methods measured on a 16 core Intel Xeon(R) W-1270P CPU. Note that MMFF, Gasteiger, and DASH do not make use of multiple cores. Models and data are preloaded into memory for all comparisons}
    \label{tab:computation_time}
\end{table}

\section{Conclusion}
In this work, a new approach to assign atomic partial charges in molecules was developed using a dynamic attention-based substructure hierarchy (DASH) in a tree structure, where the attention values are extracted from a GNN trained on high-quality QM reference charges. DASH was found to provide a prediction accuracy that is comparable to the GNN, but is independent of fast-changing and quickly deprecated ML libraries (the only requirement is basic functionality in the RDKit), directly human interpretable, and allows the retrieval of meaningful error bars on the predicted partial charges. Furthermore, assignments can be changed by the user if needed for a specific application.

To train the model, a data set was built from four different sources with a total of 393'692 unique molecules and up to three conformers per molecule. The molecules were selected to represent the substructures (as defined by Morgan fingerprints with radius 2) of the lead-like molecules in ChEMBL. The QM reference partial charges were calculated with TPSSh/def2-TZVP in an implicit solvent (dielectric permittivity $\epsilon$ of 4) and extracted with the MBIS method. These attention values from the GNN that was trained on this data set were used to order the atom types and build the DASH tree structure, where the maximal depth of the tree and the attention threshold were optimized hyperparameters. Post-assignment normalization and symmetrization ensure physically reasonable partial charges. The DASH approach outperforms commonly used methods for classical force fields such as AM1-BCC or RESP in assignment speed by two or more orders of magnitude, while predicting partial charges close to the MBIS reference. 

In this work, the DASH tree was built with MBIS as the reference charge extraction method due its high accuracy and low conformational dependency. However, the same procedure could be applied with any other type of partial charge as reference, or also for other atomic properties as target.

In conclusion, DASH is a robust, fast, and accurate method for partial charge assignments, where all assignments can be visualized as fragments of a molecule for full human readability of each partial charge assignment. The DASH tree structure and underlying source code as well as an OpenFF plug-in are freely available.


\section*{Data and Software Availability}

The implementation of the DASH-tree and all software used for this work is open source and available on GitHub (\url{https://github.com/rinikerlab/DASH-tree}). The full data set of 1'076'252 3D structures is freely available in the ETH Research Collection (DOI: 10.3929/ethz-b-000613415).

\section*{Acknowledgments}
The numerical simulations were performed on the Euler cluster operated by the High Performance Computing group at ETH Zurich. The authors gratefully acknowledge financial support by the Swiss National Science Foundation (Grant No. 200021-212732) and ETH Zurich (Grant No. ETH-50 21-1).

\printbibliography

\end{document}